\documentclass[aps,prb,twocolumn,showpacs,showkeys]{revtex4-1}
\usepackage[dvips]{graphicx}

\begin{document}

\title{On the Spectroscopic Method of Measuring the Size of the Semiconductor Nanocrystals}

\author{I.V.~Beloussov}

\email[ ]{igor.beloussov@phys.asm.md}

\author{V.~I.~Pavlenko}

\author{I.~I.~Dobinda}

\affiliation{Institute of Applied Physics, Academy of Sciences of
Moldova, 5 Academy str., Kishinev, MD-2028, Republic of Moldova }

\date{\today}

\begin{abstract}
The dependences of the fundamental transition on the semiconductor quantum dot size obtained experimentally at various temperatures using different measuring methods are analyzed and compared. The possibility to extrapolate the results for the case of arbitrary temperature is discussed.
\end{abstract}



\maketitle

Semiconductor low-dimensional systems and semiconductor quantum dots in particular, attract a considerable interest due to the substantial size dependence of the energy of quantum transitions in them. This is promising for development of absorption and luminescent materials necessary for solution of wide range of applied problems from optoelectronic imaging and data transfer devices to biological fluorescence labels.\cite{Gaponenko2005,S.Nizamoglu2007}

Owing to the size quantization of the electron and hole states in quantum dots the location of optical transitions depends on the nanocrystal radius. This phenomenon was observed experimentally when absorption spectra of semiconductor microcrystals dispergated in transparent dielectric matrices were studied.\cite{A.I.Ekimov1985,A.I.Ekimov1985a,A.I.Ekimov1989}  With the aim to explain the experimental results a theory was developed in Refs.~\onlinecite{Al.L.Efros1982,Xia1989,M.G.Bawendi1990}, which described the phenomena observed in semiconductors in the framework of a model of two simple direct zones of electrons and holes  (with the effective masses  $m_{e}$ and $m_{h}$, respectively) with a parabolic dispersion dependence. The proposed theory was constructed in the framework of the effective mass approximation, i.e., it was assumed, that all essential lengths were large relative to the lattice constant. It was also assumed that the potential well confining the motion of electrons and holes in the quantum dot possesses spherical symmetry and infinitely high walls. Moreover, it was shown that in the case of a strong dimensional quantization when the quantum dot radius $a$  is considerably greater than the Bohr radii $a_{e}$  and  $a_{h}$ of the electron and hole, respectively, the electron and hole energy levels are defined by the expression
$E_{l,n}^{e,h}=\hbar ^{2}k_{l,n}^{2}/2m_{e,h}$, where $k_{l,n}=\varphi _{l,n}/a$, and $\varphi _{l,n}$ is a universal set of numbers, which do not depend on  $a$. In the special case $l=0$  we obtain  $\varphi _{l,n}=n\pi $ ($n=1,2,3,\ldots $) and, hence, $E_{l,n}^{e,h}=\hbar ^{2}\pi ^{2}n^{2}/2m_{e,h}a^{2}\propto a^{-2}$. Since for small $a$  the distance between the energy levels is large, the electron--hole Coulomb interaction in the first order approximation was neglected.

It was shown that due to the dimensional quantization of the electron and hole levels a series of discrete lines should be observed in the interband light absorption where $\hbar \omega _{0,1}=E_{g}+\hbar ^{2}\pi ^{2}/2\mu a^{2}$  is the absorption threshold value. Here $E_{g}$  is the forbidden gap of the bulk semiconductor and $\mu =m_{e}m_{h}/\left( m_{e}+m_{h}\right) $  is the reduced mass. The complex structure of the valence band and nonparabolicity of the conduction band in semiconductor quantum dots were taken into account in Refs.~\onlinecite{A.I.Ekimov1993,S.W.Koch1992,D.J.Norris1996,Al.L.Efros1998,Al.L.Efros2000,J.Xia1999,G.B.Grigorian1990} where the effective mass approximation was also used. It was demonstrated that in this case the transitions forbidden in the framework of a simple parabolic model appear.

The electron--hole Coulomb interaction for  $a_{h}<a<a_{e}$ ($a_{e,h}=\hbar ^{2}\varepsilon /m_{e,h}e^{2}$  are the Bohr radii of the electron and hole, respectively; $\varepsilon$  is the dielectric constant, and $e$  is the electron charge) was taken into account in the adiabatic approximation.\cite{Al.L.Efros1982}  It was assumed that the energy of the electron motion considerably exceeds the energy of motion of the heavy hole; therefore, the electronic potential acting on the hole can be considered as averaged over the electron motion. It was shown that if the Coulomb interaction is taken into account, each line in the absorption spectrum transforms into a series of close lines. In this case the frequency of the fundamental transition  $\omega _{0,1}$ acquires a negative correction $\propto a^{-1}$ and a positive correction  $\propto a^{-3/2}$. Therefore, the Coulomb interaction becomes notable for quantum dots with small radii. The endeavors were made in Refs.~\onlinecite{D.B.TranThoai1990,E.Rabani1999,A.Franceschetti1997,U.E.H.Laheld1997,J.Li2000,P.E.Lippens1989,Takagahara1993,Lin-WangWang1996} and other works to take into account the Coulomb interaction of the electron and hole in detail.

In several of the aforementioned papers (for example, in Refs.~\onlinecite{A.I.Ekimov1993,P.E.Lippens1989}) the finiteness of the potential barrier, which confines the motion of the electron and hole in the quantum dot, was taken into account. Here the barrier height served as an adjustable parameter for a better compliance of the theoretical dependences and experimental data. Since real quantum dots are always surrounded by a dielectric medium, the polarization boundary effects were also taken into consideration.\cite{Y.Z.Hu1990,L.Banyai1992}

Despite many experimental results related to the optical properties of semiconductor quantum dots were qualitatively explained, the theory for small size dots is in general developed worse than for dots with greater dimensions. In particular, such an important theoretical result as the dependence of the fundamental optical transition frequency on the dot radius $a$  does not sufficiently well coincide with the experimentally obtained dependence and is usually fitted to it. This is mainly associated with the necessity to use for small $a$  concepts and methods of solid state physics, molecular physics, and quantum chemistry simultaneously (see, for example, \cite{G.Pellegrini2005} and references therein). Therefore, hereinafter while discussing this dependence we will use only the experimental data obtained in various works.

In Fig.~1 the dependences of the energy $E_{0}\left[ eV\right] $  of the fundamental optical transition (the energy of the first excited state $1S_{3/2}1S_{e}$) on the parameter $x=10^{4}/a^{2}\left[ \mathring{A}^{-2}\right] $  for CdSe quantum dots are depicted on the basis of the experimental data published in Refs.~\onlinecite{D.J.Norris1996,Lin-WangWang1996,G.Pellegrini2005,W.WilliamYu2003,D.J.Norris1994,S.Baskoutas2006,Gruenberg1997,C.B.Murray1993,Klimov2000,Klimov2002}.
\begin{figure}[!ht]
\label{fig1}
\par
\begin{center}
\includegraphics[scale=1.1]{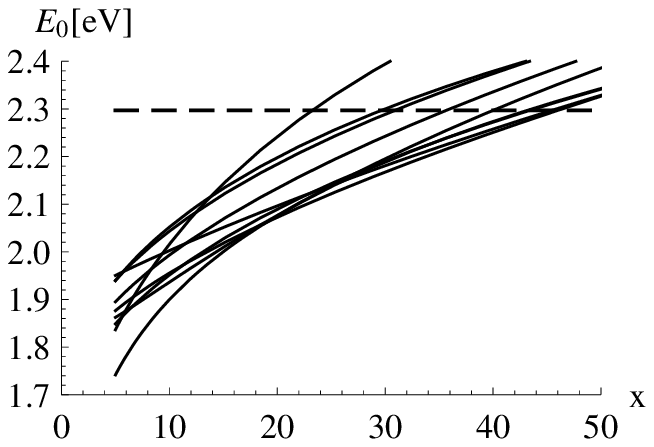}
\end{center}
\caption{Dependences of the main transition energy $E_{0}$  vs. the CdSe quantum dot radius $a$  taken from Refs.~\onlinecite{D.J.Norris1996,Lin-WangWang1996,G.Pellegrini2005,W.WilliamYu2003,D.J.Norris1994,S.Baskoutas2006,Gruenberg1997,C.B.Murray1993,Klimov2000,Klimov2002}. The horizontal dotted line corresponds to the frequency of second harmonic of YAP-laser.}
\end{figure}

At a first glance it seems that various experimental results suggest us substantially different values of the quantum dot radius for the same quantum transition and, therefore, the universal dependence of the fundamental transition energy on the radius does not exist! However, one should take into consideration that in Refs.~\onlinecite{D.J.Norris1996,D.J.Norris1994,Gruenberg1997} the results are provided based on the experiments performed at a temperature of 10 K though in Refs.~\onlinecite{Lin-WangWang1996,G.Pellegrini2005,S.Baskoutas2006,Klimov2002} on the studies performed at 300 K.\cite{W.WilliamYu2003,C.B.Murray1993,Klimov2000} The forbidden gap width and other parameters of bulk CdSe are different for various temperature values. The radii of quantum dots could be determined with insufficient accuracy. Moreover, quantum dots in various works were studied in different surrounding. Respectively, both the potential barrier height, which confines the motion of the electron and hole, and the polarization phenomena at the quantum dot surface could be different.

Another reason the mismatch of the curves can arise is that the graphical presentation of the experimental results is not sufficiently accurate. Actually, for example, the same experimental data are discussed in Ref.~\onlinecite{Gruenberg1997} and Ref.~\onlinecite{D.J.Norris1994}. However, the data of these two papers presented at the same plot as $E_{0}\left( x\right) $  dependence provide two different curves!

To be sure that the universal  $E_{0}\left( x\right) $ dependence actually exists we will do the following. First among all the curves presented in Fig. 1 we will choose one curve denoted as a reference one. We will approximate this dependence by a certain not very complicated analytical expression  $E_{0}\left( x\right) $. Each of the remaining curves we will also approximate by its own  $E_{0}\left( x\right) $ function. Then we will introduce into the resulted expressions additional parameters, which correspond to the shift of the entire curve, its extension or compression. These parameters will be chosen so as to maximally approach all the curves to the reference one. If we succeed to do so, the universal dependence obviously exists; moreover, the character of deformation of the curves itself will, probably, help us to understand, why the curves initially did not coincide.

Let as discuss separately the experimental data for CdSe quantum dots, which refer to the temperature 10K (Refs.~\onlinecite{D.J.Norris1996,D.J.Norris1994}) and 300 K (Refs.~\onlinecite{W.WilliamYu2003,C.B.Murray1993,Klimov2000}). The dependences at 10 K interesting for us are presented in Fig.~2.
\begin{figure}[!ht]
\label{fig2}
\par
\begin{center}
\includegraphics[scale=0.85]{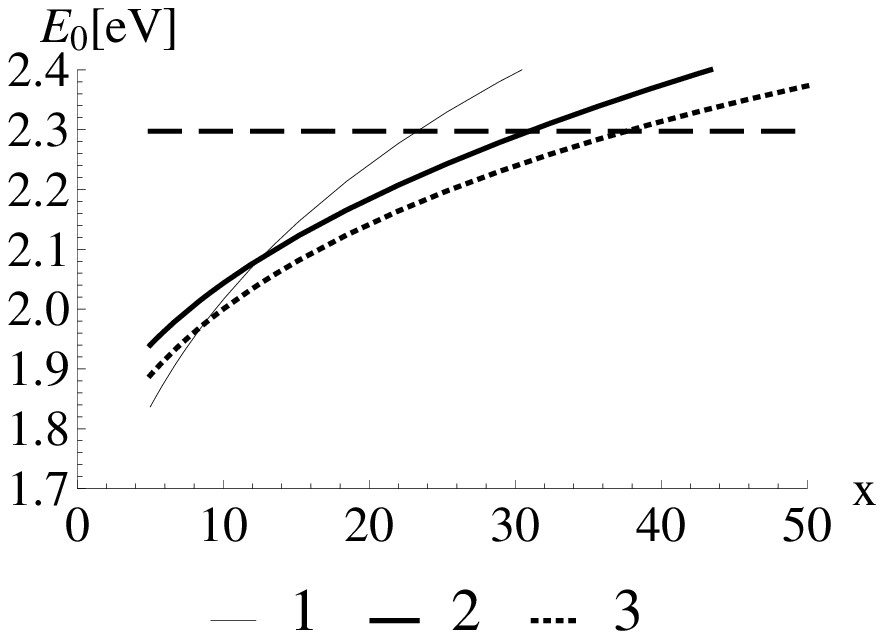}
\end{center}
\caption{Thin solid line 1 is the result fitted from Ref.~\onlinecite{D.J.Norris1996}. It depicts the dependence of the energy $E_{0}^{\left( 1\right) }\left( x\right) $  of the main transition vs. the CdSe quantum dot radius  $a$. The thick solid line 2 is the $E_{0}^{\left( 2\right) }\left( x\right) $  dependence of the first maxima of the absorption spectra vs. the radius presented in Fig.~1 of Ref.~\onlinecite{D.J.Norris1994}. The thick dashed line 3 shows the dependence $E_{0}^{\left( 3\right) }\left( x\right) $  of the pump energies vs. the radius derived from Fig.~2 of Ref.~\onlinecite{D.J.Norris1994}. }
\end{figure}

The $E_{0}^{\left( 1\right) }\left( x\right) $  dependence is obtained according the experimental data.\cite{D.J.Norris1996} Initially we tried to draw two curves in one plot. The first curve was taken from Fig.~6 of Ref.~\onlinecite{D.J.Norris1996}, the second was drawn from Figs.~1--3 of the same paper for the strongly limited quantity of data. It was found that the energy mismatch of the curves amounted to $\sim 10\div 80$~meV. This difference defines the accuracy of graphical presentation of the experimental data in the plots in Ref.~\onlinecite{D.J.Norris1996}.

We can see in Fig.~2 that though CdSe quantum dots were synthesized and characterized by the same research team (Refs.~\onlinecite{D.J.Norris1996,D.J.Norris1994}), the  $E_{0}^{\left( 1\right) }\left( x\right) $,  $E_{0}^{\left( 2\right) }\left( x\right) $, and  $E_{0}^{\left( 3\right) }\left( x\right) $ dependences are substantially different. It is easy to understand the difference between  $E_{0}^{\left( 1\right) }\left( x\right) $  and $E_{0}^{\left( 2\right) }\left( x\right) $  if to take into consideration that various measurement techniques were used in Ref.~\onlinecite{D.J.Norris1996} and Ref.~\onlinecite{D.J.Norris1994}. In Ref.~\onlinecite{D.J.Norris1996} the data were obtained using photoluminescence excitation spectroscopy (PLE), and they were used to draw the $E_{0}^{\left( 1\right) }\left( x\right) $  dependence. The  $E_{0}^{\left( 2\right) }\left( x\right) $  curve was obtained using the absorption spectra (Fig.~1 in Ref.~\onlinecite{D.J.Norris1994} or Fig.~8 in Ref.~\onlinecite{Al.L.Efros1996}).

The $E_{0}^{\left( 3\right) }\left( x\right) $  dependence is plotted versus the laser pump energy (pump-probe experiment), as noted in Ref.~\onlinecite{D.J.Norris1994} (Fig.~2 caption). To each value of the mean dot radius $a$  an energy value corresponds for which a "hole" is bleached at the absorption edge of a weak signal.

It is easy to see that the $E_{0}^{\left( 1\right) }\left( x/4\right) $  curve with high accuracy coincides with the $E_{0}^{\left( 3\right) }\left( x\right) $  curve if it is vertically shifted for a certain distance. This unexpected coincidence suggests that though everywhere in Ref.~\onlinecite{D.J.Norris1996} the concept of quantum dot radius $a$  is used, the dot's diameter $d=2a$  is actually assumed there. If this is so, the $E_{0}^{\left( 1\right) }\left( x\right) $  dependence versus the radius in Ref.~\onlinecite{D.J.Norris1996} should be represented as the curve shown as a dotted line in Fig.~2.

With the aim to determine the dependence of the fundamental transition energy versus the mean radius one should use the curve $E_{0}^{\left( 2\right) }\left( x\right) $  obtained using the absorption spectra in Refs.~\onlinecite{D.J.Norris1994,Al.L.Efros1996}. It is shifted upwards form the curve $E_{0}^{\left( 3\right) }\left( x\right) $  by the value of  $\sim 30$~meV. Note that in Ref.~\onlinecite{Gruenberg1997}, though the author cited the data from Ref.~\onlinecite{D.J.Norris1994}, he fits his theoretical dependence to the curve  $E_{0}^{\left( 3\right) }\left( x\right) $, but actually he should compare it with the curve  $E_{0}^{\left( 2\right) }\left( x\right) $.

The dependences of the fundamental transition energy on the dot radius (300 K) are presented in Fig.~3, where we also show the $E_{0}^{\left( 2\right) }\left( x\right) $  curve from Fig.~2 for comparison.
\begin{figure}[!ht]
\label{fig3}
\par
\begin{center}
\includegraphics[scale=0.85]{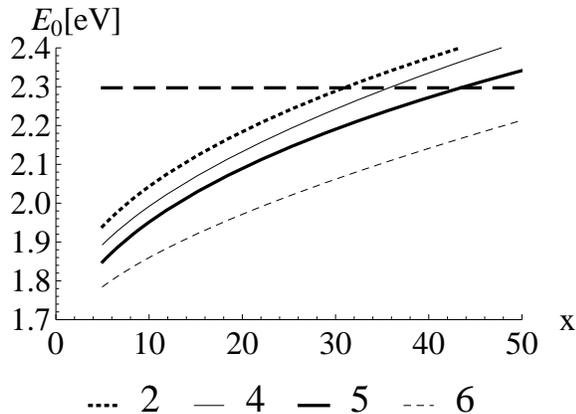}
\end{center}
\caption{Dependences of the fundamental transition energy vs. the mean quantum dot radius at 300 K: from Refs.~\onlinecite{Klimov2000,Klimov2002} (thin solid line 4), Ref.~\onlinecite{C.B.Murray1993} (thick solid line 5), and Ref.~\onlinecite{W.WilliamYu2003} (dashed line 6). The dotted line corresponds to the curve $E_{0}^{\left( 2\right) }\left( x\right) $  from Fig.~2. }
\end{figure}
\\ \

The $E_{0}^{\left( 2\right) }\left( x\right) $  and $E_{0}^{\left( 5\right) }\left( x\right) $  dependences follow from the experimental data of the same research team (Refs.~\onlinecite{D.J.Norris1994,C.B.Murray1993,Al.L.Efros1996}). The first curve was obtained at 10 K, when the bulk CdSe forbidden gap amounts to  $E_g =1.84$~eV; the second curve - at 300 K, when  $E_g =1.74$~eV. Note, that within a wide range of parameter $x$  values an approximate equation $E_{0}^{\left( 5\right) }\left( x\right)\approx E_{0}^{\left( 2\right) }\left( x\right)-1.84+1.75$
is valid. A small divergence occurs only for large $x$  values, probably, due to the conduction band nonparabolicity. The noted equation means that many spectroscopic results obtained at a certain temperature can be extrapolated to the case of other temperature values simply by taking into consideration the temperature dependence of the forbidden gap of the bulk semiconductor.

Let us transform the curves $E_{0}^{\left( 4\right) }\left( x\right) $  and  $E_{0}^{\left( 6\right) }\left( x\right) $ is such a way that they are maximally close to each other and to the curve  $E_{0}^{\left( 5\right) }\left( x\right) $. It is easy to see that for not very large $x$  values we have  $E_{0}^{\left( 5\right) }\left( x\right) \leq  E_{0}^{\left(
4\right) }\left( x\right) -0.05\approx E_{0}^{\left( 6\right) }\left(
1.3x\right) +0.05$. The transformed curves are presented in Fig.~4.
\begin{figure}[!ht]
\label{fig4}
\par
\begin{center}
\includegraphics[scale=0.85]{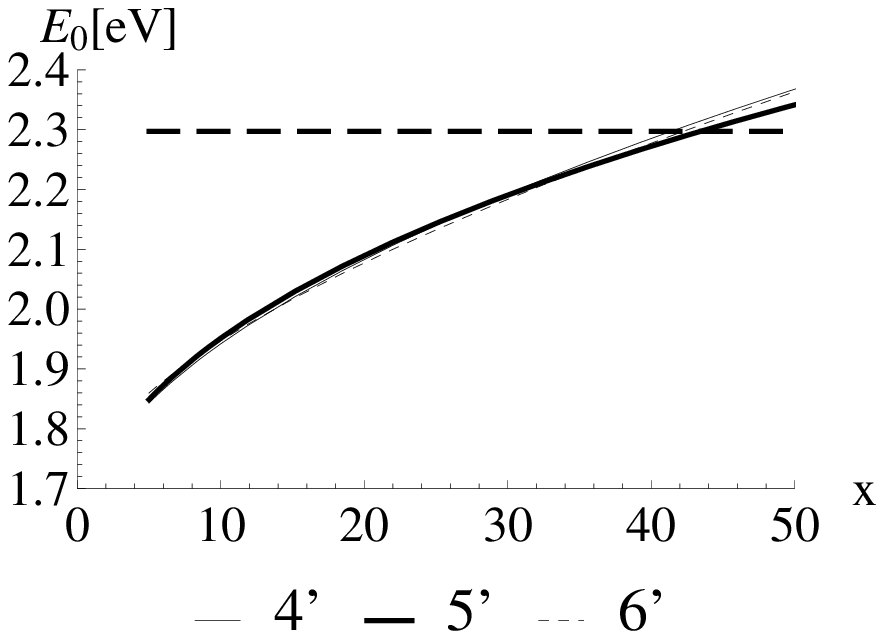}
\end{center}
\caption{
The transformed curves  $E_{0}^{\left( 4^{\prime }\right) }\left( x\right) =E_{0}^{\left( 4\right) }\left( x\right)-0.05$  (thin solid line), $E_{0}^{\left( 5^{\prime }\right) }\left( x\right) =E_{0}^{\left( 5\right) }\left( x\right)$  (thick solid line), and  $E_{0}^{\left( 6^{\prime }\right) }\left( x\right) =E_{0}^{\left( 6\right) }\left( 1.3 x\right)+0.05$ (dashed line).}
\end{figure}

It follows from Fig.~4 that the curves  $E_{0}^{\left( 4\right) }\left( x\right) $ and $E_{0}^{\left( 6\right) }\left( 1.3 x\right) $  have the same shape. Moreover, $E_{0}^{\left( 4^{\prime }\right) }\left( x\right) $  and $E_{0}^{\left( 6^{\prime }\right) }\left( x\right) $  are close to  $E_{0}^{\left( 5^{\prime }\right) }\left( x\right) $. Therefore, the divergence of the results of the works Refs.~\onlinecite{Klimov2000,Klimov2002} and Ref.~\onlinecite{W.WilliamYu2003} is probably associated with the fact that the radius of quantum dots in Ref.~\onlinecite{W.WilliamYu2003} by a factor of $\sqrt{1/1.3}\approx 0.9$  differs from the value given by the authors. If to evaluate the quantum dot size, for which the fundamental transition is, for example, in resonance with the second harmonic of YAP-laser, we obtain that for each of the curves   $E_{0}^{\left( 4^{\prime }\right) }\left( x\right) $,   $E_{0}^{\left( 5^{\prime }\right) }\left( x\right) $, and  $E_{0}^{\left( 4^{\prime }\right) }\left( x\right) $  the $a$  value divergence is within 0.4 \AA.



\end{document}